# Modified Partition Functions, Consistent Anomalies and Consistent Schwinger Terms

Amir Abbass Varshovi

amirabbassv@ipm.ir

*School of Physics, Institute for Research in Fundamental Science (IPM)*

*P. O. Box 19395-5531, Tehran, IRAN*

*Department of Physics, Sharif University of Technology*

*P. O. Box 11365-9161, Tehran, IRAN*

**Abstract:** A gauge invariant partition function is defined for gauge theories which leads to the standard quantization. It is shown that the descent equations and consequently the consistent anomalies and Schwinger terms can be extracted from this modified partition function naturally.

1. ## Introduction

In the early years of 1980s, Fujikawa showed that the chiral, the non-Abelian [1, 2], the conformal [3] and eventually the gravitational [4] anomalies can be rigorously extracted from the Jacobian matrices of relevant symmetric transformations of the Grassmannian path integral measure. In the other words, it was shown that the anomalous behaviors in quantum corrections due to the path integral quantization were in complete agreement with those derived based on the second quantization method. Specially, the calculated terms for chiral [5, 6] and non-Abelian [7, 8] anomalies in perturbation theory, coincided precisely with the Fujikawa's [1, 2] results. The cornerstone of Fujikawa's method was to Euclideanize the space-time manifold and using the celebrated index theorem [9, 10]. The crucial point was to introduce compatible regularizing exponential functions which regard the intrinsic symmetries of the theory. This idea had been firstly applied to prove the fascinating features of the Atiyah-Singer index theorem in the special case of Dirac operators (see [11, 12] and the references therein). This approach is usually called the heat kernel or the zeta-function regularization.

In fact, studying the chiral anomalies yield spectacular application of the most sophisticated topics of differential geometry in the framework of theoretical physics. Actually, the paradigm of this application was first found by Nielsen and his collaborators [13, 14] in order to derive the difference of numbers of left- and right-handed solutions of Dirac equation. The coincidence of the heat kernel regularization results and the Feynman diagram calculations in the setting of perturbation theory ensured the validity of this paradigm.

The Fujikawa's approach [1, 2], was built over a number of drastic assumptions. Initially he supposed that the functional measure is Lorentz invariant,

$$D\psi D\bar{\psi} := \prod_{x \in \mathbb{R}^4} \mathrm{d}\psi(x)\mathrm{d}\bar{\psi}(x).$$

(1)

Indeed he refused the axial invariant measure,

$$D\psi D\psi^\dagger := \prod_{x \in \mathbb{R}^4} \mathrm{d}\psi(x)\mathrm{d}\psi^\dagger(x).$$

(2)

Next, he worked out the Dirac operator as a self adjoint elliptic differential operator over $\mathbb{R}^4$ or its one-point compactification $S^4$, as a 4-dimensional spin manifold, after the redefinition; $x^0 = ix^4$. Finally he introduced a gauge invariant heat kernel to regularize the divergent Jacobian matrix determinant.

This heat kernel regularization rigorously produces the explicit form of chiral, non-Abelian and singlet anomalies in complete agreement with perturbation theory. Furthermore, the other forms of anomalies can be extracted from this method with modifying the regularizing heat kernels in appropriate settings [15-21]. As the most important case the consistent anomalies [22] can be derived from the heat kernels by replacing the ordinary Dirac operator with one of its modified counterparts [16].

In this article, it is given an alternative approach based on functional integrals to derive the explicit form of consistent anomalies for gauge theories. Moreover, this approach enables one to extract the Stora-Zumino descent equations from the path integral setting. This produces the whole anomalous terms of a quantized gauge theory including the consistent Schwinger terms in a path integral formulation. The idea is to find a modified form of partition function for gauge theories, which is substantially gauge invariant. While the different choices of integrands lead to different heat kernels in the Fujikawa's approach [15-21], this idea makes the achievements basically independent from the integrands. More precisely, the anomalous terms extracted from this modified partition function are naturally defined provided the gauge invariance is a substantial property of the modified partition function and not an imposed symmetry through the regularization programs. Furthermore, the idea of modified partition functions lets one to get rid of the choice between two equivalent functional measures (1) and (2) provided all our constructions are referred to gauge symmetry rather than any chosen Grassmannian measure which has no importance here. On the other hand, following the idea of axially extended gauge theories [23], these modified partition functions are invariant under axial transformations, which reveal that the chiral anomalies are not entirely provided by classical axial symmetry breaking. This reminds that both the vector and the axial Ward-Takahashi identities can be precisely hold by using appropriate counter terms in renormalizing the theory [15, 22]. In fact, in the method of modified partition functions, the intrinsic quantization spirit of the anomalous behaviors of an even dimensional gauge theory is clearer than that in the Fujikawa's method.

Although, in the method of modified partition functions one refuses using the heat kernels to extract the anomalous terms, but on the other hand, the appearance of the Chern classes in the formulation of modified partition functions can be considered as an evidence for the role of index theorem in anomalous behaviors.

The basic idea of modified partition functions is based on studying the possible forms of variations of the quantum action in terms of regularizing heat kernels. To see this more precisely, consider an axially extended gauge theory [23] and then choose the axial invariant functional measure (2) to define the partition function with

$$Z(A) = \int D\psi D\psi^\dagger e^{i\int_{\mathbb{R}^4} \bar{\psi} i \gamma^\mu \partial_\mu \psi + \bar{\psi} \gamma^\mu A_\mu \psi} = e^{iW(A)}.$$

(3)

Here $W(A)$ is the quantum action or the free energy. Now, consider the orbit of 1-parameter group of extended gauge transformations $e^{is\alpha} \in \mathcal{G}, s \in \mathbb{R}$, over the Affine space of the gauge fields $\mathcal{A}$ [23]. Next, suppose that the vector field of these given integral curves are denoted by $\eta^\alpha$. Recall that the BRST operator $\delta$, is the exterior derivative over the fibers of $\mathcal{A}$ [23]. Using the partition function (3) one finds that;

$$\eta^\alpha{}_A (\ln Z) = i\eta^\alpha{}_A(W) = i\delta W(\eta^\alpha{}_A) = \frac{d}{ds}\bigg|_{s=0} \det(e^{-is\alpha}) \det(e^{is\alpha}),$$

(4)

for $A \in \mathcal{A}$. The left hand side is generally non-trivial while the right hand side vanishes classically. It also seems that the right hand side needs no heat kernel regularizations. Actually, it is seen that the functional determinant could not be generally calculated in the ordinary way. On the other hand, this obstruction similarly holds for standard Yang-Mills theories. Thus, the standard results of the analytical measure theory may lead to contradictions for functional measures. This shows the need to have an alternative setting for (3).

Actually introducing the modified partition function is an attempt to understand and use the path integral quantization more mathematically. Fortunately, this attempt leads to a more consistent framework for calculating the anomalous behaviors of a gauge theory in the setting of path integrals. There still may be questions in the mathematics of modified partition functions method, but it seems that this method comes more natural and leads to a more consistent framework of path integral calculations.

In section 2, Modified Partition Function, a gauge invariant version of partition function is worked out for gauge theories based on a number of natural axioms. In section 3, the consistent anomalies and consistent Schwinger terms are extracted from the modified partition functions. It is seen that the descent equations appear in this formalism naturally. Moreover, the (anti-)BRST transformations come to the game automatically.

## 2. Modified Partition Function

In this section, a gauge invariant form of partition function for gauge theories is proposed and it is shown that a minimal set of natural axioms produces an elaborate setting to work out the explicit formula of gauge invariant partition functions in a concrete manner. This gauge invariant partition function is conventionally referred to as *modified partition function*.

Consider a given gauge theory over the space-time $\mathbb{R}^{2n}$ equipped with the Minkowskian metric. To work out the explicit formula of modified partition function for the given gauge theory, initially it is assumed that it would be of the form

$$Z_M(A) = Z(A) e^{-i \int_{\mathbb{R}^d} \Omega_d(A)}$$

(5)

for $Z(A)$ the standard partition function for the given gauge theory and for a differential $d$-form $\Omega_d$, $d \in \mathbb{N}$, with condition $\Omega_d(0) = 0$. On the other hand, the gauge field $A$ is allowed to live over $\mathbb{R}^D$, $D \geq \max\{d, 2n\}$, possibly with extra dimension components.

It is claimed that the set of axioms listed below determines the possible forms of $\Omega_d(A)$.

- a) Coherency axiom;

$$\frac{\delta^m}{\delta A^{a_m}_{\mu_m}(x_n) \ldots \delta A^{a_1}_{\mu_1}(x_1)} \ln Z_M(A) = \frac{\delta^m}{\delta A^{a_m}_{\mu_m}(x_n) \ldots \delta A^{a_1}_{\mu_1}(x_1)} \ln Z(A),$$
(6)

for any $m \in \mathbb{N}$.

- b) Relativity axiom;

$$Z_M(\mathrm{U}(\Lambda)A) = Z_M(A) \quad \text{and} \quad Z(\mathrm{U}(\Lambda)A) = Z(A)$$
(7)

for $\Lambda$ a Poincare transformation on $\mathbb{R}^{2n}$.
- c) Gauge Invariance axiom;

$$Z_M(A \triangleleft g) = Z_M(A)$$
(8)

for any gauge transformation $g \in \mathcal{G}$ (see [23]).
- d) Flatness axiom;

$\Omega_d(A)|_U = 0$ if and only if the connection is flat over open set $U \subseteq \mathbb{R}^D$.

(9)

Here to formulate the coherency axiom one should use the Frechet derivative more precisely to define the functional derivatives in (6) in a stricter setting. Hence suppose that $(M, \mu_M)$ and $(N, \mu_N)$ are measure smooth manifolds [24, 25] with Lebesgue measures $\mu_M$ and $\mu_N$ and $N \subseteq M$ is a closed embedded submanifold. This leads to a family of deformation retracts for topological spaces of test functions over $M$ and $N$; $C_c^\infty(N) \xrightarrow{i_n} C_c^\infty(M) \xrightarrow{r} C_c^\infty(N)$, $1 \leq n$, where $i_n$ is the inclusion map with $\mu_M(\mathrm{supp}\, i_n(g) \setminus N) \leq \frac{1}{n}$ for any $g \in C_c^\infty(N)$, while the projection $r$ is the ordinary restriction map.

Now consider a family of distributions, $\Lambda^M{}_n$, $n \in \mathbb{N}$, defined with compactly supported smooth functions $\delta f_n$ by $\Lambda^M{}_n g = \int_M \delta f_n\, g\, d\mu_M$, $g \in C_c^\infty(M)$, with $\lim_{n\to\infty} \Lambda^M{}_n = \hat{\delta}_M(m)$, $m \in M$, for $\hat{\delta}_M$ the Dirac distribution on $M$. Here we use $\hat{\delta}_M$ for the Dirac distribution and $\delta_M$ for the Dirac delta function over $M$. The ordinary functional derivation $\frac{\delta}{\delta f(m)} \int_M \mathcal{F}(f)\, d\mu_M = \int_M \frac{\delta}{\delta f(m)} \mathcal{F}(f)\, d\mu_M$ for $f \in C_0^\infty(M)$, $m \in M$, and for $\mathcal{F}(z)$ a harmonic function on $\mathbb{C}$ is defined by;

$$\lim_{n\to\infty} \int_M \frac{d}{d\varepsilon}\bigg|_{\varepsilon=0} \mathcal{F}(f + \varepsilon \delta f_n)\, d\mu_M = \lim_{n\to\infty} \Lambda^M{}_n \mathcal{F}'(f) = \mathcal{F}'(f)(m).$$

(10)

Thus, (10) is usually written by; $\frac{\delta}{\delta f(m)} \mathcal{F}(f) = \mathcal{F}'(f)\, \delta_M(m)$ which leads to definition $\frac{\delta}{\delta f(m)} f := \delta_M(m)$ as the functional derivation with respect to $f(m)$. More strictly, the functional derivation with respect to $f(m)$ through $N$ is defined as a distribution on $N$ given by; $\lim_{n_1 \to \infty} \lim_{n_2 \to \infty} i_{n_2}{}^*(\Lambda^M{}_{n_1})$. Thus;

$$\frac{\delta}{\delta_N f(m)} f = \delta_{\dim N}^{\dim M}\, \delta_M(m) = \begin{cases} 0 & , \dim N < \dim M \\ \delta_M(m) & , \dim N = \dim M \end{cases},$$

(11)

for $m \in M$. Conversely, for a given $g \in C_0^\infty(N)$, the functional derivative with respect to $g(m)$ through $M$ is a distribution on $M$ defined by; $\lim_{n\to\infty} r^*(\Lambda^N{}_n)$. But since $\lim_{n\to\infty} r^*(\Lambda^N{}_n) = \hat{\delta}_M$ one finds that; $\frac{\delta}{\delta_M g(m)} g = \delta_M(m)$, $m \in N$. For simplicity, for any $f \in C_0^\infty(M)$, the functional derivative with respect to $f(m)$ through $M$, is denoted by the standard notation $\frac{\delta}{\delta f(m)}$. Moreover, when $N = \mathbb{R}^k$, $\frac{\delta}{\delta_N f(m)}$ and $\delta_N$ are simply denoted by $\frac{\delta}{\delta_k f(m)}$ and $\delta^{(k)}$, respectively.

Thus by definitions above and for strictly speaking, the functional derivatives in (6) all have to be replaced by the functional derivatives through $\mathbb{R}^{2n}$.

Now let us describe the given four axioms more precisely. According to axiom (a), the $n$-point functions defined by the modified partition function are precisely the standard $n$-point functions. Therefore, modified partition functions lead to the standard quantization of gauge theories. This is the reason for using the terminology of Coherency axiom. The condition of $\Omega_d(0) = 0$ implies that $Z_M(0) = Z(0)$, which can be considered as the weak Coherency axiom. The non-modified part of the axiom (b) insists on the Poincare invariance of the fermion path integral measure, i.e.;

$$D\psi D\psi^\dagger = D\psi D\bar\psi = DU(\Lambda)\psi D\bar\psi U(\Lambda)^{-1} = DU(\Lambda)\psi\ D\psi^\dagger U(\Lambda)^\dagger,$$

(12)

for any Poincare transformation $\Lambda$. On the other hand, the axiom (b) asserts that $\Omega_d(A)$ is Poincare invariant, since the theory is substantially relativistic. Axiom (c) is the expected property for the partition function of a gauge theory. In the case of axial gauge theories [23], this axiom contains all the extended gauge transformations. Eventually, as it is shown in the following, the last axiom helps one to find the possible forms of $\Omega_d$. Actually the axiom of Flatness asserts that $\Omega_d$, as a differential $d$-form, is a function of the curvature $F = dA + A^2$ only.

Axiom (a) implies that the following relation should be hold;

$$\int_{\mathbb{R}^d} \frac{\delta}{\delta_{2n} A_\mu^a(x)} \Omega_d(A) = \int_{\mathbb{R}^d} \delta_{2n}^d\ \delta^{(d)}(x) \times \ldots = 0.$$

(13)

The dotted term ... in the integral is zero if and only if $\Omega_d(A)$ is itself equal to zero. Therefore; $d > 2n$. This shows that $\Omega_d(A)$ cannot be a local term as we expected before, since otherwise a local term could compensate the classical symmetry breakings due to the anomalies. It is not true, since if it was the case, an appropriate set of renormalizing counter terms could remove the anomalies completely.

For the axiom (b), consider a 1-parameter group of Poincare transformations $X_t$, $t \in \mathbb{R}$, and the induced curves on the Affine space of the gauge fields $\mathcal{A}$ by $X_t^*(A)$. The axiom of Relativity asserts that; $Z_M(X_t^*(A)) = Z_M(A)$, and thus; $\frac{d}{dt}\big|_{t=0} Z_M(X_t^*(A)) = 0$. Hence (12) leads to;

$$\frac{d}{dt}\bigg|_{t=0} X_t^*\big(\Omega_d(A)\big) = L_X\big(\Omega_d(A)\big) = 0,$$

(14)

Thus, for any translation generator $X \in \mathbb{R}^{2n}$, $L_X\big(\Omega_d(A)\big) = 0$. On the other hand, it is obvious that translating along the extra dimensions should keep $\Omega_d(A)$ invariant, since otherwise a nonphysical translation would lead to physical observations through the modified partition function. Thus, for any translation generator $X \in \mathbb{R}^D$, $L_X\big(\Omega_d(A)\big) = 0$. Moreover, $\Omega \in \Omega_{\text{deR}}^d(\mathbb{R}^D)$ is closed if and only if $L_X \Omega = 0$ for any translation generator $X \in \mathbb{R}^D$. Therefore;

$$d\Omega_d(A) = 0$$

(15)

for any $A$. By Poincare's lemma there exists a $d-1$-form, say $\Omega_{d-1}(A)$, such that;

$$\Omega_d(A) = d\Omega_{d-1}(A).$$

(16)

Now without losing generality we replace $\mathbb{R}^d$ by $\mathbb{R}^{d-1} \times \mathbb{R}_{\geq 0}$ in (5). Hence, since $\partial(\mathbb{R}^{d-1} \times \mathbb{R}_{\geq 0}) = \mathbb{R}^{d-1}$, by Stokes' theorem we have;

$$Z_M(A) = Z(A) e^{-i \int_{\mathbb{R}^{d-1}} \Omega_{d-1}(A)}$$

(17)

Then, the Flatness axiom can be modified to: $\Omega_{d-1}(A)|_U = 0$ if and only if $F = dA + A^2 = 0$ over open set $U \subseteq \mathbb{R}^D$. Moreover, using the Coherency axiom and (13) again, one finds that; $d - 1 > 2n$.

Now it is the time to use the axiom of Gauge Invariance. This axiom implies that for a 1-parameter group of gauge transformations, say $e^{ti\alpha}$, $t \in \mathbb{R}$, one should have $\frac{d}{dt}\big|_{t=0} Z_M(A \triangleleft e^{ti\alpha}) = 0$. Therefore;

$$\int_{\mathbb{R}^{2n}} \alpha^a D^a(J) - \int_{\mathbb{R}^{d-1}} \delta\Omega_{d-1}(\eta^\alpha{}_A) = 0 ,$$

(18)

where $\delta$ is the BRST operator and $\eta^\alpha$ is the vector field of (4). Taking the functional derivative $\delta/\delta_{2n}\alpha^a(x)$ of (18), one finds that;

$$D^a(J)(x) - \int_{\mathbb{R}^{d-1}} \delta_{2n}^{d-1} \delta^{(d-1)}(x) \times \ldots = 0 .$$

(19)

For axial gauge theories including chiral ones, $D^a(J)(x)$ do not vanish generally, but (19) with $d - 1 > 2n$ asserts that $D^a(J)(x) = 0$ unless in (18), $\delta\Omega_{d-1}(A)$ be an exact deRham form ($\delta\Omega_{d-1}(A) = d\Omega_{d-2}^1(A)$), $\mathbb{R}^{d-1}$ be replaced by $\mathbb{R}^{d-2} \times \mathbb{R}_{\geq 0}$ and eventually $d$ be equal to $2n + 2$. The anti-commutation relation of d and $\delta$ implies that; $\delta\Omega_d(A) = \delta d\Omega_{d-1}(A) = -d^2\delta\Omega_{d-2}^1(A)$. Hence, we find that;

$$\delta\Omega_{2n+2}(A) = 0 .$$

(20)

Therefore, if there exists a solution of equations (9), (15) and (20) for a given gauge theory, then it admits a gauge invariant partition function. In the following, it is seen that different solutions for $\Omega_{2n+2}(A)$ leads to different anomalies and Schwinger terms. Thus, it seems that different solutions for $\Omega_{2n+2}(A)$ correspond to different renormalization methods.

The only if part of the Flatness axiom lets one to write down $\Omega_{2n+2}(A)$ only in terms of the curvature $F = dA + A^2$. Gauge invariance implies that $\Omega_{2n+2}(A)$ must be an invariant polynomial [26]. Moreover, since the Newton's formula expresses all the invariant forms by trace, then any linear combination of differential forms

$$\prod_{i=1}^{m} Tr\{F^{k_i}\}$$

(21)

with $\sum_{i=1}^{m} k_i = n + 1$ can be considered as a solution for $\Omega_{2n+2}(A)$. Thus, the general solution is given in terms of the *Chern characters*;

$$Tr\{e^F\} = \sum_{k=0}^{\infty} \frac{1}{k!} Tr\{F^k\}.$$

(22)

Therefore, for 2 and 4-dimensional non-Abelian $SU(N)$-gauge theories in fundamental representations, the only possible solutions for $\Omega_d(A)$ are given in terms of $\Omega_4(A) = Tr\{F^2\}$ and $\Omega_6(A) = Tr\{F^3\}$ respectively, the suggested terms in [27-29]. But for higher dimensional $SU(N)$-gauge theories more than one solution exist. Despite of non-Abelian gauge theories, the Abelian ones have unique solutions for $\Omega_{2n+2}(A)$ (up to a constant factor) in all even dimensions by $\Omega_{2n+2}(A) = F^{n+1}$. Thus, modified partition functions could be uniquely defined for 2 and 4-dimensional non-Abelian $SU(N)$-gauge theories and for all even dimensional Abelian ones.

## 3. Consistent Anomalies and Consistent Schwinger Terms

In this section, the explicit form of consistent anomalies and consistent Schwinger terms are derived for a given gauge theory using the idea of the modified partition functions. The consistent anomaly can be extracted from the Fujikawa's approach when one replaces the ordinary Dirac operator with $P_+\gamma^\mu D_\mu + P_-\gamma^\mu \partial_\mu$ for $P_\pm = (\frac{1\pm\gamma_5}{2})$ [16]. But the idea of modified partition functions enables one to extract the consistent anomalies for the standard Yang-Mills theories. Moreover, this idea also enables one to calculate the consistent Schwinger terms for the currents algebra.

As mentioned in section 2, one can consider the following modified partition function for a $2n$-dimensional gauge theory;

$$Z_M(A) = Z(A)e^{-i\int_{\mathbb{R}^{2n}\times\mathbb{R}_{\geq 0}} c_n \Omega_{2n+1}(A)},$$

(23)

where $Tr\{F^{n+1}\} = d\Omega_{2n+1}(A)$ and $c_n \in \mathbb{C}$ is a dimension dependant factor which is given by the index theorem. Consider the 1-parameter group gauge transformation $e^{ti\alpha} \in \mathcal{G}$ and its relevant vector field $\eta^\alpha$. Gauge invariance implies that $\frac{d}{dt}\big|_{t=0} Z_M(A \triangleleft e^{ti\alpha}) = 0$, and thus one concludes that;

$$\int_{\mathbb{R}^{2n}} \{\alpha^a D^a(J) - c_n \Omega^1_{2n}(A)(\eta_\alpha)\} = 0,$$

(24)

where $\delta\Omega_{2n+1}(A) = d\Omega^1_{2n}(A)$. The functional derivation $\frac{\delta}{\delta\alpha^a(x)}$ gives rise to the well-known formula for the consistent anomaly;

$$D^a(J)(x) = c_n(\Omega^1_{2n}(A))^a(x),$$

(25)

where $(\Omega^1_{2n}(A))^a$ is defined by;

$$\Omega^1_{2n}(A) = \omega^a(\Omega^1_{2n}(A))^a,$$

(26)

for colored ghosts $\omega^a$. Note that it can be supposed that the gauge theory is automatically extended in the sense of [23] and thus $\omega^a$s can be both ghosts and anti-ghosts. Therefore, using the idea of modified partition functions to extract the gauge anomalies, the crucial role of axial symmetry in anomalous behaviors is diminished. Moreover, since the consistent anomalous terms defined by Stora-Zumino descent equations [27-29] have cohomological interpretations [27, 30, 31, 32, 33], we conclude that modified partition functions lead to a cohomological description of path integral quantization.

Gauge invariance asserts that $Z_M(A)$ is constant over the fibers of the Affine space of all gauge fields $\mathcal{A}$ modulo gauge transformations [23]. Actually, the variation of the effective action $W(A)$ within the fibers can be calculated through the gauge transformation loops $g: S^1 \to \mathcal{G}$ [16]. This variation also can be evaluated in the context of noncommutative geometry, using the local index formula [34-36]. In fact, $Z(A)$ is a continuous function and thus by (3) the variation of $W(A)$ should be equal to $2\pi m$ for some $m \in \mathbb{Z}$ determined by the index theorem [16].

Now it is the time to show that the consistent Schwinger terms can also be extracted from modified partition functions. Conventionally, the Schwinger terms are defined by;

$$[J^{a0}(t,x), J^{b0}(t,y)]\Big|_{anomalous} = i\mathcal{S}c^{ab}(t,x) \times \delta^{(2n-1)}(x-y).$$

(27)

for $J^{a0}$, the time component of colored current $J^a$.

Now set

$$\Xi_{2n}^{(t_0;x,y)}(\varepsilon_1, \varepsilon_2)$$

$$= i\frac{\delta}{\delta_{2n} A_0^a(t_0+\varepsilon_1, x)} i\frac{\delta}{\delta_{2n} A_0^b(t_0-\varepsilon_2, y)} - i\frac{\delta}{\delta_{2n} A_0^a(t_0-\varepsilon_2, x)} i\frac{\delta}{\delta_{2n} A_0^b(t_0+\varepsilon_1, y)}$$

(28)

as a functional differential operator and assume that;

$$\lim_{\varepsilon \to 0} \varepsilon\, \delta(\varepsilon) = 1$$

(29)

for $\delta(\varepsilon) = \delta^{(1)}(\varepsilon)$, the 1-dimensional Dirac delta function. Moreover, to regularize the singular terms in the following, some achievements of measure theory should be used [24, 25]. In fact, it is known that for any locally integrable function $f \in L^1_{loc}(\mathbb{R}^k)$, the following equality holds for almost every $x \in \mathbb{R}^k$;

$$\lim_{r \to 0} \frac{1}{\mu(B(x;r))} \int_{B(x;r)} f\, d\mu = f(x),$$

(30)

where $B(x;r) \subseteq \mathbb{R}^k$ is a ball centered at $x$ with radius $r$ and $\mu$ is the Lebesgue measure. Here it is supposed that for any function $f: \mathbb{R}_{\geq 0} \times \mathbb{R}_{\geq 0} \to \mathbb{C}$ in below, the following version of theorem (30) holds;

$$f(0) = \lim_{\varepsilon_0 \to 0^+} \frac{1}{\varepsilon_0^2} \int_{[0,\varepsilon_0] \times [0,\varepsilon_0]} f.$$

(31)

Obviously, (31) holds for any continuous function over $\mathbb{R}_{\geq 0} \times \mathbb{R}_{\geq 0}$, but in the following, (31) is used to regularize the bad behaved functions at their singular points. Actually (31) is a rough understanding of the well-known Lebesgue differentiation theorem [25].

To extract the consistent Schwinger term, note that the smoothness of $Z_M(A)$ leads to;

$$\lim_{\varepsilon_1, \varepsilon_1 \to 0^+} \Xi_{2n}^{(t_0;x,y)}(\varepsilon_1, \varepsilon_2)\, Z_M(A) = 0.$$

(32)

Changing the variables $t = t_0 - \varepsilon_2$ and $\varepsilon = \varepsilon_1 + \varepsilon_2$ results in;

$$\lim_{\varepsilon \to 0^+} \{i \frac{\delta}{\delta_{2n} A_0^a(t+\varepsilon, x)} i \frac{\delta}{\delta_{2n} A_0^b(t, y)} - i \frac{\delta}{\delta_{2n} A_0^a(t, x)} i \frac{\delta}{\delta_{2n} A_0^b(t+\varepsilon, y)}\} \ln Z_M(A)$$

$$= \lim_{\varepsilon \to 0^+} \{i \frac{\delta}{\delta_{2n} A_0^a(t+\varepsilon, x)} i \frac{\delta}{\delta_{2n} A_0^b(t, y)} - i \frac{\delta}{\delta_{2n} A_0^a(t, x)} i \frac{\delta}{\delta_{2n} A_0^b(t+\varepsilon, y)}\} \ln Z(A)$$

$$-i \lim_{\varepsilon \to 0^+} \{i \frac{\delta}{\delta_{2n} A_0^a(t+\varepsilon, x)} i \frac{\delta}{\delta_{2n} A_0^b(t, y)} - i \frac{\delta}{\delta_{2n} A_0^a(t, x)} i \frac{\delta}{\delta_{2n} A_0^b(t+\varepsilon, y)}\} \int c_n \Omega_{2n+1}(A)$$

$$= 0 .$$

(33)

The former term is the standard formula;

$$\lim_{\varepsilon \to 0^+} \{i \frac{\delta}{\delta_{2n} A_0^a(t+\varepsilon, x)} i \frac{\delta}{\delta_{2n} A_0^b(t, y)} - i \frac{\delta}{\delta_{2n} A_0^a(t, x)} i \frac{\delta}{\delta_{2n} A_0^b(t+\varepsilon, y)}\} \ln Z(A)$$

$$= \langle [J^{a0}(t, x), J^{b0}(t, y)] \rangle .$$

(34)

Therefore, (33) leads to;

$$\langle [J^{a0}(t, x), J^{b0}(t, y)] \rangle$$

$$= -i \lim_{\varepsilon \to 0^+} \{\frac{\delta}{\delta_{2n} A_0^a(t+\varepsilon, x)} \frac{\delta}{\delta_{2n} A_0^b(t, y)} - \frac{\delta}{\delta_{2n} A_0^a(t, x)} \frac{\delta}{\delta_{2n} A_0^b(t+\varepsilon, y)}\} \int_{z \in \mathbb{R}^{2n} \times \mathbb{R}_{\geq 0}} c_n \Omega_{2n+1}(A)(z)$$

$$= -i c_n \lim_{\varepsilon \to 0^+} \int_{z_1, z_2 \in \mathbb{R}^{2n}, z \in \mathbb{R}^{2n} \times \mathbb{R}_{\geq 0}} \{(\delta_{z_1}^{(2n)}(t+\varepsilon, x) \delta_{z_2}^{(2n)}(t, y) - \delta_{z_1}^{(2n)}(t, x) \delta_{z_2}^{(2n)}(t+\varepsilon, y))$$

$$\frac{\delta}{\delta_{2n} A_0^a(z_1)} \frac{\delta}{\delta_{2n} A_0^b(z_2)} \Omega_{2n+1}(A)(z)\} .$$

(35)

Now set;

$$\delta_{z_1}^{(2n)}(t+\varepsilon,x)\delta_{z_2}^{(2n)}(t,y) - \delta_{z_1}^{(2n)}(t,x)\delta_{z_2}^{(2n)}(t+\varepsilon,y)$$

$$= \delta_{z_1}^{(2n)}(t+\varepsilon,x)\delta_{z_2}^{(2n)}(t,y) - \delta_{z_1}^{(2n)}(t,x)\delta_{z_2}^{(2n)}(t,y)$$

$$+\delta_{z_1}^{(2n)}(t,x)\delta_{z_2}^{(2n)}(t,y) - \delta_{z_1}^{(2n)}(t,x)\delta_{z_2}^{(2n)}(t+\varepsilon,y)$$

$$= \varepsilon\delta_{z_1}^{(2n)}(t,x)\partial_0^{z_2}\delta_{z_2}^{(2n)}(t+\varepsilon,y) - \varepsilon\partial_0^{z_1}\delta_{z_1}^{(2n)}(t+\varepsilon,x)\delta_{z_2}^{2n}(t,y).$$

(36)

Note that the derivations should be worked out at $t+\varepsilon$. Specially, to calculate the equal-time commutators, one should replace $\mathbb{R}^{2n}$ by $\mathbb{R}^{2n-1}\times\mathbb{R}_{\geq 0}$ with its boundary located at $z^0 = t$, and work out the derivations at $z^0 = t+\varepsilon$ in the body of the manifold. Thus, (35) up to factor $-ic_n$, equals to;

$$\lim_{\varepsilon\to 0^+} \varepsilon\int_{z_1,z_2\in\mathbb{R}^{2n},z\in\mathbb{R}^{2n}\times\mathbb{R}_{\geq 0}} \{(\delta_{z_1}^{(2n)}(t,x)\partial_0^{z_2}\delta_{z_2}^{(2n)}(t+\varepsilon,y) - \partial_0^{z_1}\delta_{z_1}^{(2n)}(t+\varepsilon,x)\delta_{z_2}^{(2n)}(t,y))$$

$$\frac{\delta}{\delta_{2n}A_0^a(z_1)}\frac{\delta}{\delta_{2n}A_0^b(z_2)}\Omega_{2n+1}(A)(z)\}.$$

(37)

Using the gauge transformation $e^{ti\alpha}$ and its induced vector field $\eta^\alpha$, one finds;

$$\int_{z_1\in\mathbb{R}^{2n},z\in\mathbb{R}^{2n}\times\mathbb{R}_{\geq 0}} \partial_0^{z_1}\delta_{z_1}^{(2n)}(t+\varepsilon,x)\frac{\delta}{\delta_{2n}A_0^a(z_1)}\Omega_{2n+1}(A)(z)$$

$$= \int_{z_1\in\mathbb{R}^{2n},z\in\mathbb{R}^{2n}\times\mathbb{R}_{\geq 0}} -\partial_i^{z_1}\delta_{z_1}^{(2n)}(t+\varepsilon,x)\frac{\delta}{\delta_{2n}A_i^a(z_1)}\Omega_{2n+1}(A)(z)$$

$$+C^{afg}\int_{z\in\mathbb{R}^{2n}\times\mathbb{R}_{\geq 0}} A_\mu^f(t+\varepsilon,x)\frac{\delta}{\delta_{2n}A_\mu^g(t,x)}\Omega_{2n+1}(A)(z)$$

$$-\frac{\delta}{\delta_{2n}\alpha^a(t+\varepsilon,x)}\delta\Omega_{2n+1}(\eta^\alpha{}_A)$$

(38)

for $\delta$ the BRST derivation and for $\mu = 0, \ldots, D-1$ and $i = 1, \ldots, D-1$, while $C^{afg}$ are the structure constants of the gauge group. The same equation can be derived similarly for the other term including $\partial_0^{z_2} \delta_{z_2}^{(2n)}(t, y) \frac{\delta}{\delta_{2n} A_0^b(z_2)}$. To this end, consider the gauge transformation $e^{ti\beta}$ and its induced vector field $\eta^\beta$. The result is the same as (38) but with replacing $\alpha$, $a$, $z_1$ and $x$ by $\beta$, $b$, $z_2$ and $y$ respectively. The descent equation; $\delta \Omega_{2n+1}(A) = d\Omega_{2n}^1(A)$, together with (38) make (37) to be;

$$\lim_{\varepsilon \to 0^+} \varepsilon \Big\{ \int_{z_2 \in \mathbb{R}^{2n}, z \in \mathbb{R}^{2n}} \delta_{z_2}^{(2n)}(t, y) \frac{\delta}{\delta A_0^b(z_2)} \frac{\delta}{\delta \alpha^a(t+\varepsilon, x)} \Omega_{2n}^1(\eta^\alpha{}_A)(z)$$

$$- \int_{z_1 \in \mathbb{R}^{2n}, z \in \mathbb{R}^{2n}} \delta_{z_1}^{(2n)}(t, x) \frac{\delta}{\delta A_0^a(z_1)} \frac{\delta}{\delta \beta^b(t+\varepsilon, y)} \Omega_{2n}^1(\eta^\beta{}_A)(z)$$

$$+ C^{bfg} \int_{z_1 \in \mathbb{R}^{2n}, z \in \mathbb{R}^{2n} \times \mathbb{R}_{\geq 0}} \delta_{z_1}^{(2n)}(t, x) \frac{\delta}{\delta_{2n} A_0^a(z_1)} A_\mu^f(t+\varepsilon, y) \frac{\delta}{\delta_{2n} A_\mu^g(t, y)} \Omega_{2n+1}(A)(z)$$

$$- C^{afg} \int_{z_2 \in \mathbb{R}^{2n}, z \in \mathbb{R}^{2n} \times \mathbb{R}_{\geq 0}} \delta_{z_2}^{(2n)}(t, y) \frac{\delta}{\delta_{2n} A_0^b(z_2)} A_\mu^f(t+\varepsilon, x) \frac{\delta}{\delta_{2n} A_\mu^g(t, x)} \Omega_{2n+1}(A)(z)$$

$$+ \int_{z_1, z_2 \in \mathbb{R}^{2n}, z \in \mathbb{R}^{2n} \times \mathbb{R}_{\geq 0}} \partial_i^{z_1} \delta_{z_1}^{(2n)}(t+\varepsilon, x) \delta_{z_2}^{(2n)}(t, y) \frac{\delta}{\delta_{2n} A_i^a(z_1)} \frac{\delta}{\delta_{2n} A_0^b(z_2)} \Omega_{2n+1}(A)(z)$$

$$- \int_{z_1, z_2 \in \mathbb{R}^{2n}, z \in \mathbb{R}^{2n} \times \mathbb{R}_{\geq 0}} \delta_{z_1}^{(2n)}(t, x) \partial_i^{z_2} \delta_{z_2}^{(2n)}(t+\varepsilon, y) \frac{\delta}{\delta_{2n} A_0^a(z_1)} \frac{\delta}{\delta_{2n} A_i^b(z_2)} \Omega_{2n+1}(A)(z).$$

(39)

Direct calculations show that all the terms of (39) vanish according to (11) and the limit $\varepsilon \to 0$ except the two first ones. Now set;

$$\delta_{z_i}^{(2n)}(t, y) = \delta_{z_i}^{(2n)}(t+\varepsilon, y) + \varepsilon \partial_0^{z_i} \delta_{z_i}^{(2n)}(t+\varepsilon, y)$$

(40)

for $i = 1, 2$. Thus, (39) would be of the following form;

$$\lim_{\varepsilon \to 0^+} \varepsilon \{ \int_{z_2 \in \mathbb{R}^{2n}, z \in \mathbb{R}^{2n}} \delta_{z_2}^{(2n)}(t+\varepsilon, y) \frac{\delta}{\delta A_0^b(z_2)} \frac{\delta}{\delta \alpha^a(t+\varepsilon, x)} \Omega_{2n}^1(\eta^\alpha{}_A)(z)$$

$$- \int_{z_1 \in \mathbb{R}^{2n}, z \in \mathbb{R}^{2n}} \delta_{z_1}^{(2n)}(t+\varepsilon, x) \frac{\delta}{\delta A_0^a(z_1)} \frac{\delta}{\delta \beta^b(t+\varepsilon, y)} \Omega_{2n}^1(\eta^\beta{}_A)(z) \}$$

$$\lim_{\varepsilon \to 0^+} \varepsilon^2 \{ \int_{z_2 \in \mathbb{R}^{2n}, z \in \mathbb{R}^{2n}} \partial_0^{z_2} \delta_{z_2}^{(2n)}(t+\varepsilon, y) \frac{\delta}{\delta A_0^b(z_2)} \frac{\delta}{\delta \alpha^a(t+\varepsilon, x)} \Omega_{2n}^1(\eta^\alpha{}_A)(z)$$

$$- \int_{z_1 \in \mathbb{R}^{2n}, z \in \mathbb{R}^{2n}} \partial_0^{z_1} \delta_{z_1}^{(2n)}(t+\varepsilon, x) \frac{\delta}{\delta A_0^a(z_1)} \frac{\delta}{\delta \beta^b(t+\varepsilon, y)} \Omega_{2n}^1(\eta^\beta{}_A)(z) \}$$

(41)

It can be checked that the first term takes the form of;

$$\lim_{\varepsilon \to 0^+} \varepsilon \int_{z \in \mathbb{R}^{2n}} \delta_z^{(2n)}(t+\varepsilon, x) \delta_z^{(2n)}(t+\varepsilon, y) \times \dots$$

(42)

for a generally non-vanishing term .... Thus (42) is proportional to $\lim_{\varepsilon \to 0^+} \varepsilon \, \delta(0)$ which is obviously divergent. Considering an uncertainty in time proportional to $\varepsilon^2$, makes (42) be proportional to

$$\lim_{\varepsilon \to 0^+} \varepsilon \, \delta(\mathcal{O}(\varepsilon^2)),$$

(43)

which is still singular according to (29). To regularize it by (31) note that;

$$\int_{\varepsilon_2=0}^{\varepsilon_0} \int_{\varepsilon_1=0}^{\varepsilon_0} (\varepsilon_1 + \varepsilon_2) \, \delta((\varepsilon_1 + \varepsilon_2)^2) \, d\varepsilon_1 d\varepsilon_2 = 0,$$

(44)

which cancels out the first term of (41). The second term of (41) can also be rewritten in the sense of (38). Therefore, following the same approach gives (41) in the form of;

$$\lim_{\varepsilon \to 0^+} \varepsilon^2 \int_{z \in \mathbb{R}^{2n}} \frac{\delta}{\delta \alpha^a(t+\varepsilon, x)} \frac{\delta}{\delta \beta^b(t+\varepsilon, y)}$$

$$\{\delta\left(\Omega^1_{2n}(\eta^\beta)\right)(\eta^\alpha{}_A)(z) - \delta(\Omega^1_{2n}(\eta^\alpha))(\eta^\beta{}_A)(z)\}$$

$$+ \dots .$$

(45)

Similarly it can be shown that the dotted term vanishes according to (31). Now recall that; $d\omega(X_0, X_1) = X_0\omega(X_1) - X_1\omega(X_0) - \omega([X_0, X_1])$ for any 1-form $\omega$ and two vector fields $X_0$ and $X_1$. Considering the BRST operator as the exterior derivative in the sense of [23], one concludes that;

$$\delta\left(\Omega^1_{2n}(\eta^\beta)\right)(\eta^\alpha{}_A) - \delta(\Omega^1_{2n}(\eta^\alpha))(\eta^\beta{}_A)$$

$$= \Omega^1_{2n}\left([\eta^\alpha, \eta^\beta]_A\right) + \delta\Omega^1_{2n}(\eta^\alpha{}_A, \eta^\beta{}_A),$$

(46)

for $\eta^\alpha$ and $\eta^\beta$ the vector fields over $\mathcal{A}$ induced by gauge transformations $e^{ti\alpha}$ and $e^{ti\beta}$ respectively. Note that $e^{ti\alpha}$ and $e^{ti\beta}$ can be considered as extended gauge transformations for axially extended gauge theories [23]. However, the first term of (46) results in the canonical term while the second term leads to the anomalous one. To calculate the anomalous term, one should consider the standard descent equation; $\delta\Omega^1_{2n}(A) = d\Omega^2_{2n-1}(A)$. Moreover, as mentioned above $\mathbb{R}^{2n}$ should be replaced by $\mathbb{R}^{2n-1} \times \mathbb{R}_{\geq 0}$, while the boundary is considered to be at $z^0 = t$. This restricts the integral in (45) to $\mathbb{R}^{2n-1}$. On the other hand, the functional derivatives in (45) are taken through $\mathbb{R}^{2n}$ which lead to $2n$-dimensional Dirac delta functions $\delta^{(2n)}$. Therefore, one concludes that;

$$\langle[J^{a\,0}(t,x),J^{b\,0}(t,y)]\rangle\Big|_{anomalous}$$

$$= -ic_n \lim_{\varepsilon\to 0} \varepsilon^2 \delta(\varepsilon)\delta(\varepsilon) \int_{z\in\mathbb{R}^{2n-1}} \delta_z^{(2n-1)}(x)\delta_z^{(2n-1)}(y)(\Omega^2_{2n-1}(A))^{a\,b}(z)$$

$$= -ic_n \lim_{\varepsilon\to 0} \varepsilon^2 \delta(\varepsilon)\delta(\varepsilon)\delta^{(2n-1)}(x-y)(\Omega^2_{2n-1}(A))^{a\,b}(t,x),$$

(47)

where $(\Omega^2_{2n}(A))^{a\,b}$s are defined by;

$$\Omega^2_{2n}(A) = \omega^a \omega^b (\Omega^2_{2n-1}(A))^{a\,b},$$

(48)

with $\omega^a$s the colored ghosts. Actually, for an axially extended gauge theory, $\omega^a$s can be simultaneously ghosts and anti-ghosts. Indeed, from (47), one derives the ghost/anti-ghost and the anti-ghost/anti-ghost Schwinger terms in addition to the ordinary ghost/ghost Schwinger terms [23].

By (29), (47) leads to;

$$\langle[J^{a\,0}(t,x),J^{b\,0}(t,y)]\rangle\Big|_{anomalous}$$

$$= -ic_n \delta^{(2n-1)}(x-y)(\Omega^2_{2n-1}(A))^{a\,b}(t,x).$$

(49)

Comparing with (27), one finds that

$$\mathcal{S}c^{a\,b}(t,x) = -c_n (\Omega^2_{2n-1}(A))^{a\,b}(t,x),$$

(50)

as was expected before. Indeed, (50) is in complete agreement with the well-known Mickelsson-Faddeev-Shatashvili cocycle for the quantized currents algebra [37-39].

This shows that the modification of partition functions enables one to non-perturbatively extract the Schwinger terms from the path integral formalism. Consequently, by (25) and (49), it is shown that the idea of modified partition functions give more fascinating features to the relation of path integral quantization and anomalous behaviors of gauge theories which lead to such spectacular achievements.

Finally, it would be beneficial to briefly discuss the canonical term of (46). The second quantization asserts that;

$$\langle [J^{a^0}(t,x), J^{b^0}(t,y)] \rangle \Big|_{canonical}$$
$$= iC^{abc}\delta^{(2n-1)}(x-y)\langle J^{c^0}(t,x) \rangle,$$

(51)

while from (45) and (46) one concludes that;

$$\langle [J^{a^0}(t,x), J^{b^0}(t,y)] \rangle \Big|_{canonical}$$
$$= -ic_n C^{abc}\delta^{(2n-1)}(x-y) \lim_{\varepsilon \to 0} \varepsilon^2 \int_{z\in\mathbb{R}} \delta_{z^0}(t+\varepsilon)\delta_{z^0}(t+\varepsilon) \left(\Omega_{2n}^1(A)\right)^c(z,x),$$

(52)

since $[\eta^\alpha, \eta^\beta] = \eta^{[\alpha,\beta]}$. Similarly, to find the canonical term (52), one should calculate the terms of the form $\lim_{\varepsilon \to 0} \varepsilon^2 \delta(\mathcal{O}(\varepsilon^2))$ which are finite according to (29). Thus, the expectation values of time components of colored currents are intimately related to the topological factor $c_n$ and the uncertainty in time.

# Conclusion

In this article, it was shown that the standard partition functions of gauge theories can be modified to a gauge invariant formalism. Actually, considering four natural axioms (Coherency, Relativity, Gauge Invariance and Flatness), leads to a gauge invariant formulation of partition function for all even dimensional gauge theories which is referred to as modified partition function. Moreover, the modified partition functions for all even dimensional $U(1)$-gauge theories and for 2 and 4-dimensional non-Abelian $SU(N)$-gauge theories were found uniquely. Also it was seen that for higher dimensional non-Abelian gauge theories the modified partition functions have no unique formalisms but all their possible forms were derived precisely. Eventually, it was shown that modifying the partition function of a gauge theory leads to an appropriate setting to calculate the consistent anomalies and Schwinger terms from the path integral formulation. Moreover, extracting the anomalous terms from the modified partition functions gives a deeper understanding of descent equations within the path integral quantization. On the other hand, the idea of axially extended gauge theories [23] is completely compatible with the framework of modified partition functions and thus the results of [23] can be used here naturally.

# Acknowledgement

The author states his gratitude to F. Ardalan, A. Akhavan, M. Khalkhali and N. Sadooghi for comments and discussions. Moreover, my special thanks go to S. Ziaee for many things. Finally, I have to acknowledge my special gratitude to A. Shafiei Deh Abad, for his beneficial helps and guidance.

21. J. M. Gipson, *Path-Integral Derivation of Gauge and Gravitational Chiral Anomalies in Theories with Vector and Axial-Vector Couplings in Arbitrary Even Dimensions*, Phys. Rev. D 33: 1061-1078, 1986.
22. S. Weinberg, *The Quantum Theory of Fields*, vol. II, Cambridge University Press, 1996.
23. A. A. Varshovi, *Axial Symmetry, Anti-BRST Invariance and Modified Anomalies*, 32 pages [arXiv:hep-th/1011.1095].
24. W. Rudin, *Functional Analysis*, 2nd ed., McGraw-Hill, Inc. 1991.
25. G. B. Folland, *Real Annalysis, Modern Techniques and Their Applications*, 2nd ed., John Weily & Sons, Inc. 1999.
26. W. A. Poor, *Differential Geometric structures*, McGraw-Hill Book Company, 1981.
27. B. Zumino, *Cohomology of Gauge Groups: Cocycles and Schwinger Terms*, Nuc. Phys. B 253: 477-493, 1985.
28. B. Zumino, *in Relativity, Groups and Topology II*, eds. B. S. Dewitt and R. Stora (NorthHolland, Amsterdam, 1984):1293.
29. J. Manes, R. Stora and B. Zumino, *Algebraic Study of Chiral Anomalies*, Commun. Math. Phys. 102: 157-174, 1985.
30. L. D. Faddeev, *Operator Anomaly for Gauss's Law*, Phys. Lett. B 145, 81-84, 1984.
31. C. Ekstrand, *A Simple Algebraic Derivation of the Covariant Anomaly and Schwinger Term*, J. Math. Phys. 41: 7294-7303, 2000 [arXive:hep-th/9903147].
32. C. Ekstrand, *A Geometrical Description of the Consistent and Covariant Chiral Anomaly*, Phys. Lett. B 485: 193-201, 2000 [arXive:hep-th/0002238].
33. C. Adam, C. Ekstrand and T. Sykora, *Covariant Schwinger Terms*, Phys. Rev. D62, no. 105033, 15 pages, 2000 [arXive:hep-th/0005019].
34. A. Connes and H. Moscovici, *The Local Index Formula in Noncommutative Geometry*, GAFA 5: 174-243, 1995.
35. D. Perrot, *BRS Cohomology and the Chern Character in Non-Commutative Geometry*, Lett. Math. Phys. 50: 135-144, 1999 [arXive:hep-th/9910044].
36. D. Perrot, *Anomalies and Noncommutative Index Theory*, Contemp. Math. 434: 125-160, 2007 [arXive:hep-th/0603209].
37. J. Mickelsson, *On a Relation Between Massive Yang-Mills Theories and Dual String Models*, Lett. Math. Phys. 7: 45-50, 1983.
38. J. Mickelsson, *Chiral Anomalies in Even and Odd Dimensions*, Commun. Math. Phys. 97: 361-370, 1985.
24